\documentclass[a4paper,11pt]{article}
\usepackage{jcappub} 

\arxivnumber{} 
\title{\boldmath Self-Interacting Dark Matter-Induced Dynamical Friction and Its Imprint on EMRI Gravitational Waves}

\author[a,b]{YuWang}
\author[b]{Meilin Liu}
\author[a]{Haiguang Xu}

\affiliation[a]{School of Physics and Astronomy, Shanghai Jiao Tong University, Shanghai 200240, China}

\affiliation[b]{School of Aeronautics and Astronautics, Shanghai Jiao Tong University, Shanghai 200240, China}

\emailAdd{sjtu2686361@sjtu.edu.cn}
\emailAdd{meilin.liu@sjtu.edu.cn}
\emailAdd{hgxu@sjtu.edu.cn}
\date{\today}

\abstract{Extreme-mass-ratio inspirals (EMRIs) provide a promising probe of the dark
matter environment surrounding massive black holes through their long-duration
gravitational-wave signals. In this work, we investigate the influence of
self-interacting dark matter (SIDM) on EMRI dynamics and gravitational-wave
phase evolution. Motivated by the collisional Boltzmann description of SIDM,
we construct a leading-order effective correction to the conventional
Chandrasekhar dynamical-friction force, controlled by the momentum-transfer
cross section $\sigma_T$ and the dimensionless collisionality parameter
$n_\chi\sigma_T r=r/\lambda_{\rm mfp}$. We then compute the resulting orbital
evolution, waveform modification, and accumulated gravitational-wave
dephasing. In the enhanced-drag regime considered here, increasing the SIDM
self-interaction strength increases the effective dynamical friction,
accelerates the inspiral, and produces a progressively larger phase deviation
relative to the collisionless dark matter case. We further distinguish the
total dark-matter-induced dephasing relative to vacuum from the additional
phase correction generated specifically by dark matter self-interactions.
Our results demonstrate that the cumulative gravitational-wave phase of EMRIs
can be sensitive to the microscopic momentum-transfer properties of dark
matter, suggesting that future space-based detectors such as LISA may provide
a complementary probe of SIDM in the vicinity of massive black holes.
}

\begin{document}
\maketitle
\flushbottom

\section{Introduction}
\label{sec:intro}

Extreme-mass-ratio inspirals (EMRIs), consisting of a stellar-mass compact
object orbiting a massive black hole, are among the most promising
gravitational-wave (GW) sources for future space-based detectors such as the
Laser Interferometer Space Antenna (LISA)
~\cite{AmaroSeoane2017}. Unlike comparable-mass binary systems,
EMRIs evolve through a long inspiral phase during which the emitted
gravitational waves accumulate a large number of orbital cycles, typically
$\mathcal{O}(10^5)$--$\mathcal{O}(10^6)$
~\cite{BarackCutler2004,AmaroSeoane2007}. This long-term phase coherence
makes EMRIs particularly sensitive to small deviations from idealized vacuum
Kerr dynamics, including modifications of the background spacetime and
environmental perturbations
~\cite{Gair2013,Barausse2014}. EMRIs therefore provide not only a powerful
probe of strong-field gravity, but also a sensitive tool for investigating
the matter distribution and dynamical processes in the vicinity of massive
black holes.

Dark matter (DM), which accounts for most of the matter content of the
Universe~\cite{Planck2018}, may develop enhanced density distributions around
massive black holes through gravitational evolution and the adiabatic growth
of the central object. The long-term evolution of matter distributions in
galactic nuclei is strongly influenced by the gravitational potential of the
central massive black hole~\cite{Merritt2010}. In particular, adiabatic
growth can generate steep dark matter density profiles commonly referred to
as dark matter spikes~\cite{GondoloSilk1999}. These dense environments can
influence the dynamics of compact objects orbiting close to the black hole.
In addition to the energy and angular-momentum losses caused by
gravitational-wave emission, an inspiraling object moving through the dark
matter distribution experiences dynamical friction associated with the
gravitational wake generated in the surrounding medium
~\cite{Chandrasekhar1943}. This additional dissipative mechanism can modify
both the orbital evolution and the accumulated gravitational-wave phase
~\cite{Eda2013,Macedo2013,Cardoso2022,Nichols2023}.

Many previous studies of environmental effects on EMRIs have considered
collisionless dark matter and modeled the drag using the classical
Chandrasekhar prescription~\cite{BinneyTremaine2008}. A variety of dark
matter density distributions, including Navarro--Frenk--White and Einasto
halos, as well as cusp- and spike-like configurations around massive black
holes, have been widely studied in astrophysical and gravitational-wave
contexts
~\cite{Navarro1997,Einasto1965,Eda2013,Macedo2013,Cardoso2022,Wang2025CQG}.
Recent studies have further shown that microscopic dark matter interaction
processes can modify the inner structure of dark matter spikes and leave
characteristic imprints on EMRI gravitational-wave phase evolution
~\cite{Wang2025CQG,Wang2025SIDM}. More generally, gravitational-wave
observations have been proposed as a powerful probe of dark matter
distributions and particle properties~\cite{Kavanagh2017}.

However, the microscopic nature of dark matter remains unknown, and a broad
class of particle models allows for non-negligible dark matter
self-interactions. Such interactions are commonly characterized by the
momentum-transfer cross section $\sigma_T$, which quantifies the efficiency
of momentum exchange and transport in the dark matter medium
~\cite{Spergel2000,Feng2010,Petraki2014,DelNobile2015,Chu2020}.

Self-interacting dark matter (SIDM) provides a well-motivated extension of
the collisionless dark matter paradigm. Self-scattering can modify halo
structure, redistribute particle momenta, and alter the velocity distribution
and transport properties of the dark matter medium
~\cite{TulinYu2018,Kaplinghat2016,Vogelsberger2012,Rocha2013}. Recent studies
have also shown that dark matter self-interactions can significantly modify
dynamical friction and gravitational-wave-related dynamics in compact-object
systems. Numerical simulations indicate that SIDM can enhance dynamical
friction relative to collisionless dark matter in a broad range of
parameters, although reduced drag can occur in some low-velocity regimes
~\cite{FischerSagunski2024}. Self-interacting scalar dark matter has also
been shown to produce additional gravitational-wave phase corrections in
binary black hole systems~\cite{Boudon2024}, while SIDM dynamical friction
has been investigated in connection with the evolution of supermassive black
hole binaries and the final-parsec problem~\cite{AlonsoAlvarez2024}.
Recent work has further demonstrated that strongly interacting dark matter
can modify spike structures and generate distinguishable EMRI phase
signatures through different microscopic interaction channels
~\cite{Wang2025SIDM}. Nevertheless, the direct modification of the
dynamical-friction force by SIDM self-scattering, and its cumulative impact
on the long-duration phase evolution of EMRIs, remains comparatively less
explored.

The large number of accumulated cycles in an EMRI makes the
gravitational-wave phase particularly sensitive to weak dissipative
corrections. Even a small modification of the instantaneous orbital decay
rate can accumulate into an appreciable dephasing over the inspiral.
Environmental effects and other deviations from vacuum evolution can
therefore leave significant imprints on EMRI gravitational-wave signals
~\cite{Barausse2014,Cardoso2022,Berti2015,Maselli2022}. This sensitivity
motivates the use of EMRIs as a probe not only of the macroscopic dark matter
distribution, but also of its microscopic scattering properties.

In this work, we investigate the influence of SIDM self-scattering on EMRI
dynamics and gravitational-wave phase evolution. We begin with a
particle-physics-motivated scalar-mediated self-interaction model and derive
the corresponding momentum-transfer cross section. Motivated by the
collisional Boltzmann description, we then construct a leading-order
effective correction to the Chandrasekhar dynamical-friction force in terms
of the characteristic collisionality parameter
\begin{equation}
n_\chi\sigma_T r
=
\frac{r}{\lambda_{\rm mfp}}.
\end{equation}
We incorporate this effective drag into the EMRI orbital evolution and
calculate the resulting gravitational-wave frequency and phase evolution.

We focus on the enhanced-drag branch of the effective SIDM response, for
which self-scattering increases the magnitude of the dynamical friction.
Within this regime, increasing the momentum-transfer cross section accelerates
the inspiral and generates an additional accumulated gravitational-wave phase
shift relative to the collisionless dark matter case. We distinguish the
total dark-matter-induced dephasing relative to vacuum from the additional
phase contribution generated specifically by SIDM self-interactions. Our
results show that the latter can reach the order-radian level for
representative parameters in the LISA frequency band, illustrating the
potential of EMRI phase measurements to probe the microscopic transport
properties of dark matter.

The remainder of this paper is organized as follows.
In Sec.~\ref{sec:sidm_model}, we introduce the self-interacting dark matter
model and derive the corresponding momentum-transfer cross section.
In Sec.~\ref{sec:sidm_df}, we develop the SIDM-modified dynamical-friction
model motivated by the collisional Boltzmann framework.
In Sec.~\ref{sec:emri_dynamics}, we investigate the effects of
SIDM-induced dynamical friction on the orbital evolution and
gravitational-wave signals of EMRIs.
In Sec.~\ref{sec:phase_evolution}, we derive the gravitational-wave
frequency and phase evolution in the presence of SIDM-induced dynamical
friction. The numerical results and their physical implications are
presented in Sec.~\ref{sec:results}. Finally, Sec.~\ref{sec:conclusions}
summarizes our main conclusions and discusses possible directions for
future work.

\section{Dark Matter Self-Interactions}
\label{sec:sidm_model}

We consider a generic self-interacting dark matter (SIDM) particle $\chi$
with mass $m_\chi$. For definiteness, we assume that the self-interaction is
mediated by a light scalar field $\phi$ through the interaction
\begin{equation}
\mathcal{L}_{\rm int}
=
g_\chi \bar{\chi}\chi\phi ,
\end{equation}
where $g_\chi$ is the dark matter--mediator coupling and $m_\phi$ denotes
the mediator mass. Light-mediator models provide a simple and widely studied
framework for generating velocity-dependent dark matter self-interactions
~\cite{TulinYu2018,BuckleyFox2010}.

In the nonrelativistic regime relevant for astrophysical systems, scalar
exchange gives rise to an effective Yukawa potential,
\begin{equation}
V(r)
=
-\alpha_\chi
\frac{e^{-m_\phi r}}{r},
\end{equation}
where
\begin{equation}
\alpha_\chi
=
\frac{g_\chi^2}{4\pi}.
\end{equation}
Such Yukawa-type interactions can lead to strongly velocity-dependent
self-scattering, especially when the mediator is much lighter than the dark
matter particle.

The finite interaction range $m_\phi^{-1}$ leads to a velocity-dependent
self-scattering cross section. A useful dimensionless parameter is
\begin{equation}
R
=
\frac{m_\chi v_{\rm rel}}{m_\phi},
\end{equation}
which compares the characteristic momentum transfer with the mediator mass.

For astrophysical applications, the total cross section is generally less
relevant than the momentum-transfer cross section,
\begin{equation}
\sigma_T
=
\int d\Omega\,
(1-\cos\theta)
\frac{d\sigma}{d\Omega},
\end{equation}
because the factor $(1-\cos\theta)$ suppresses forward scattering and
measures the efficiency of momentum exchange between dark matter particles
.

\subsection{Self-scattering cross section}

In the Born regime, the nonrelativistic differential cross section associated
with the Yukawa interaction can be written as
\begin{equation}
\frac{d\sigma}{d\Omega}
=
\frac{\alpha_\chi^2 m_\chi^2}
{\left[
m_\phi^2
+
m_\chi^2v_{\rm rel}^2
\sin^2(\theta/2)
\right]^2},
\end{equation}
which leads to the standard velocity-dependent self-scattering behavior of
light-mediator SIDM models~\cite{TulinYu2018,Petraki2014}.

Integrating over the scattering angle yields the momentum-transfer cross
section
\begin{equation}
\sigma_T
=
\frac{8\pi\alpha_\chi^2}
{m_\chi^2v_{\rm rel}^4}
\left[
\ln(1+R^2)
-
\frac{R^2}{1+R^2}
\right],
\end{equation}
where
\begin{equation}
R
=
\frac{m_\chi v_{\rm rel}}{m_\phi}.
\end{equation}

In the short-range or low-velocity limit, $R\ll1$, the momentum-transfer
cross section approaches a constant,
\begin{equation}
\sigma_T
\simeq
\frac{4\pi\alpha_\chi^2m_\chi^2}
{m_\phi^4}.
\end{equation}

In the opposite regime, $R\gg1$, the interaction becomes effectively
long-ranged and the cross section decreases approximately as
\begin{equation}
\sigma_T
\simeq
\frac{8\pi\alpha_\chi^2}
{m_\chi^2v_{\rm rel}^4}
\left[
\ln R^2-1
\right].
\end{equation}

The resulting velocity dependence allows SIDM models to exhibit different
effective interaction strengths in environments with different characteristic
velocity dispersions. In the following, the
momentum-transfer cross section $\sigma_T$ enters the characteristic
self-scattering rate and determines the leading-order effective correction
to the dynamical-friction force acting on the inspiraling compact object.

\section{Dynamical Friction in a Self-Interacting Dark Matter Medium}
\label{sec:sidm_df}

In the collisionless limit, a compact object of mass $M$ moving with
velocity $v$ through a dark matter medium of density $\rho_\chi$
experiences the Chandrasekhar dynamical-friction force
~\cite{Chandrasekhar1943,BinneyTremaine2008},
\begin{equation}
\mathbf{F}_{\rm df}^{(0)}
=
-4\pi G^2M^2\rho_\chi
\frac{\ln\Lambda}{v^2}
\hat{\mathbf v},
\end{equation}
where $\ln\Lambda$ is the Coulomb logarithm.

\subsection{Self-interaction correction}

In a self-interacting dark matter medium, the gravitational wake generated
by the inspiraling compact object can be modified by dark matter
self-scattering~\cite{FischerSagunski2024}. The phase-space distribution of
dark matter particles is governed by the collisional Boltzmann equation
~\cite{TulinYu2018},
\begin{equation}
\frac{\partial f}{\partial t}
+
\mathbf v\cdot\nabla_{\mathbf x}f
-
\nabla_{\mathbf x}\Phi\cdot\nabla_{\mathbf v}f
=
C[f],
\end{equation}
where $f(\mathbf x,\mathbf v,t)$ is the dark matter distribution function,
$\Phi$ is the gravitational potential, and $C[f]$ denotes the collision
operator associated with dark matter self-interactions.

To connect the microscopic self-scattering process with the gravitational
wake, we write the distribution function and gravitational potential as
\begin{equation}
f=f_0+\delta f,
\qquad
\Phi=\Phi_0+\Phi_p+\Phi_{\rm w},
\end{equation}
where $f_0$ and $\Phi_0$ describe the unperturbed dark matter background,
$\Phi_p$ is the potential generated by the inspiraling compact object, and
$\Phi_{\rm w}$ denotes the gravitational potential generated by the induced
dark matter wake.

To linear order in the perturbations, the Boltzmann equation becomes
\begin{equation}
\frac{\partial \delta f}{\partial t}
+
\mathbf v\cdot\nabla_{\mathbf x}\delta f
-
\nabla_{\mathbf x}
\left(
\Phi_p+\Phi_{\rm w}
\right)
\cdot
\nabla_{\mathbf v} f_0
=
C^{(1)}[\delta f].
\end{equation}

In the collisionless limit, $C^{(1)}=0$. Introducing Fourier modes of the form
\begin{equation}
\delta f,\,
\delta\Phi
\propto
\exp\left[
i(\mathbf k\cdot\mathbf x-\omega t)
\right],
\end{equation}
the linearized kinetic equation gives
\begin{equation}
\delta f(\mathbf k,\mathbf v,\omega)
=
-
\frac{
\mathbf k\cdot\nabla_{\mathbf v}f_0
}{
\omega-\mathbf k\cdot\mathbf v+i0^+
}
\delta\Phi(\mathbf k,\omega),
\end{equation}
where
\begin{equation}
\delta\Phi
=
\Phi_p+\Phi_{\rm w}.
\end{equation}

The corresponding density perturbation is
\begin{equation}
\delta\rho
=
m_\chi
\int d^3v\,\delta f.
\end{equation}
It is therefore useful to define the collisionless gravitational
susceptibility
\begin{equation}
\chi_0(\mathbf k,\omega)
=
-
m_\chi
\int d^3v\,
\frac{
\mathbf k\cdot\nabla_{\mathbf v}f_0
}{
\omega-\mathbf k\cdot\mathbf v+i0^+
}.
\end{equation}

The wake potential is related to the density perturbation through the Poisson
equation,
\begin{equation}
-k^2\Phi_{\rm w}
=
4\pi G\,\delta\rho.
\end{equation}
Introducing the gravitational response function
\begin{equation}
\epsilon(\mathbf k,\omega)
=
1+
\frac{4\pi G}{k^2}
\chi_0(\mathbf k,\omega),
\end{equation}
the wake potential can formally be written as
\begin{equation}
\Phi_{\rm w}
=
\left[
\frac{1}{\epsilon(\mathbf k,\omega)}
-1
\right]
\Phi_p.
\end{equation}

The dynamical-friction force corresponds to the gravitational backreaction
of the wake on the compact object,
\begin{equation}
\mathbf F_{\rm df}
=
-M
\nabla\Phi_{\rm w}
\bigg|_{\mathbf x=\mathbf x_p}.
\end{equation}
For a compact object moving locally with velocity $\mathbf V$, the perturbing
potential satisfies
\begin{equation}
\omega=\mathbf k\cdot\mathbf V,
\end{equation}
and hence the drag force may be expressed formally as
\begin{equation}
\mathbf F_{\rm df}
=
-M
\int
\frac{d^3k}{(2\pi)^3}
\,i\mathbf k
\left[
\frac{1}{
\epsilon(\mathbf k,\mathbf k\cdot\mathbf V)
}
-1
\right]
\Phi_p(\mathbf k).
\end{equation}

For a Maxwellian dark matter background, the collisionless limit of this
response reproduces the conventional Chandrasekhar dynamical-friction
prescription. In the simplified form adopted in this work, it is written as
\begin{equation}
\mathbf{F}_{\rm df}^{(0)}
=
-4\pi G^2M^2\rho_\chi
\frac{\ln\Lambda}{v^2}
\hat{\mathbf v}.
\end{equation}

In the presence of self-interactions, the collision operator modifies the
phase-space response and therefore changes the structure of the gravitational
wake. For elastic self-scattering, the characteristic momentum-transfer rate
is controlled by the momentum-transfer cross section $\sigma_T$. The local
self-scattering rate may be estimated as
\begin{equation}
\Gamma_{\rm SIDM}
\simeq
n_\chi\sigma_T v,
\end{equation}
where
\begin{equation}
n_\chi
=
\frac{\rho_\chi}{m_\chi}
\end{equation}
is the dark matter number density.

The associated mean free path is
\begin{equation}
\lambda_{\rm mfp}
=
\frac{1}{n_\chi\sigma_T}.
\end{equation}

For a compact object orbiting at radius $r$, the characteristic dynamical
timescale is
\begin{equation}
t_{\rm dyn}
\sim
\frac{r}{v}.
\end{equation}
The characteristic number of self-scattering events occurring over this
timescale is therefore
\begin{equation}
N_{\rm scat}
\sim
\Gamma_{\rm SIDM}t_{\rm dyn}
\simeq
n_\chi\sigma_T r
=
\frac{r}{\lambda_{\rm mfp}}.
\end{equation}

The dimensionless quantity
\begin{equation}
\epsilon_{\rm SIDM}
\equiv
n_\chi\sigma_T r
=
\frac{r}{\lambda_{\rm mfp}}
\end{equation}
therefore provides a natural measure of the collisionality of the dark matter
medium on the characteristic scale of the gravitational wake.

In general, the full self-interacting dynamical-friction force can be regarded
as a functional of the collision operator and hence of the collisionality
parameter,
\begin{equation}
\mathbf F_{\rm df}^{\rm SIDM}
=
\mathbf F_{\rm df}
\left(
\epsilon_{\rm SIDM}
\right).
\end{equation}
In the weakly collisional regime,
\begin{equation}
\epsilon_{\rm SIDM}
=
n_\chi\sigma_T r
\ll 1,
\end{equation}
the response can be expanded perturbatively around the collisionless solution,
\begin{equation}
\mathbf F_{\rm df}^{\rm SIDM}
=
\mathbf F_{\rm df}^{(0)}
\left[
1+
\eta_{\rm eff}
\epsilon_{\rm SIDM}
+
\mathcal O
\left(
\epsilon_{\rm SIDM}^2
\right)
\right].
\end{equation}

The coefficient $\eta_{\rm eff}$ depends, in general, on the detailed form of
the collision operator, the dark matter velocity distribution, and the
structure of the gravitational wake. Rather than attempting to solve the full
collisional Boltzmann equation, we adopt a phenomenological leading-order
description and denote the effective response coefficient by $\eta$.

In the present work, we focus on the enhanced-drag branch,
\begin{equation}
\eta>0,
\end{equation}
for which self-scattering increases the effective gravitational drag acting
on the inspiraling compact object. Retaining only the leading-order
correction, we obtain
\begin{equation}
\mathbf F_{\rm df}^{\rm SIDM}
=
\mathbf F_{\rm df}^{(0)}
\left(
1+\eta n_\chi\sigma_T r
\right).
\end{equation}

For quasi-circular EMRI orbits, the orbital velocity approximately satisfies
\begin{equation}
v^2
\simeq
\frac{GM_{\rm BH}}{r},
\end{equation}
and therefore
\begin{equation}
r
=
\frac{GM_{\rm BH}}{v^2}.
\end{equation}

The self-interaction correction can thus be rewritten as
\begin{equation}
\mathbf F_{\rm df}^{\rm SIDM}
=
\mathbf F_{\rm df}^{(0)}
\left[
1+
\eta n_\chi\sigma_T
\frac{GM_{\rm BH}}{v^2}
\right].
\end{equation}

Using
\begin{equation}
n_\chi
=
\frac{\rho_\chi}{m_\chi},
\end{equation}
together with the Chandrasekhar expression, the effective dynamical-friction
force becomes
\begin{equation}
\mathbf F_{\rm df}^{\rm SIDM}
=
-4\pi G^2M^2\rho_\chi
\frac{\ln\Lambda}{v^2}
\left[
1+
\eta
\frac{\rho_\chi}{m_\chi}
\sigma_T
\frac{GM_{\rm BH}}{v^2}
\right]
\hat{\mathbf v}.
\end{equation}
The correction is therefore controlled by the dimensionless quantity
$n_\chi\sigma_T r=r/\lambda_{\rm mfp}$, which measures the importance of
self-scattering over the characteristic scale of the gravitational wake. The
above expression should be interpreted as a leading-order phenomenological
description motivated by the collisional Boltzmann framework, rather than as
an exact solution of the full self-interacting kinetic problem.

Because EMRIs accumulate a large number of orbital cycles, even a moderate
modification of the dynamical-friction force can produce a substantial
cumulative gravitational-wave phase shift. In the following, we use this
effective prescription to investigate the orbital evolution, waveform
modification, and gravitational-wave dephasing induced by dark matter
self-interactions.

\section{Effects of SIDM-Induced Dynamical Friction on EMRI Dynamics}
\label{sec:emri_dynamics}

Before turning to the quantitative gravitational-wave phase analysis, we
first illustrate the qualitative influence of SIDM-induced dynamical friction
on the orbital dynamics and waveform morphology. For this purpose, we perform
a dimensionless parameter study in geometrized units,
\begin{equation}
G=c=M_{\rm BH}=1,
\end{equation}
and vary the characteristic dark matter density parameter $\rho_0$, the
effective dissipative parameter $\mu$, and the dimensionless self-interaction
parameter $\sigma_T$.

The purpose of this calculation is to demonstrate the dynamical response of
the system to changes in the environmental and self-interaction parameters.
The parameter values adopted in this section should therefore be interpreted
as dimensionless model parameters rather than as a direct representation of
a unique astrophysical system. Physical values of the dark matter
self-interaction cross section are considered separately in the
gravitational-wave phase analysis below.

The orbital evolution is obtained by numerically integrating the Hamiltonian
equations supplemented by dissipative terms describing gravitational
radiation reaction and dark-matter-induced dynamical friction. The latter is
implemented through an effective modification of the Chandrasekhar drag,
whose generic form is
\begin{equation}
F_{\rm DF}
=
F_{\rm DF}^{(0)}
\left(
1+\delta_{\rm SI}
\right),
\end{equation}
where $F_{\rm DF}^{(0)}$ denotes the collisionless dynamical-friction
contribution and $\delta_{\rm SI}$ represents the additional correction
associated with dark matter self-interactions. In the dimensionless
parameter scan presented here, increasing $\sigma_T$ corresponds to
increasing the strength of this effective SIDM correction.

\begin{figure}[htbp]
\centering
\includegraphics[width=0.85\linewidth]{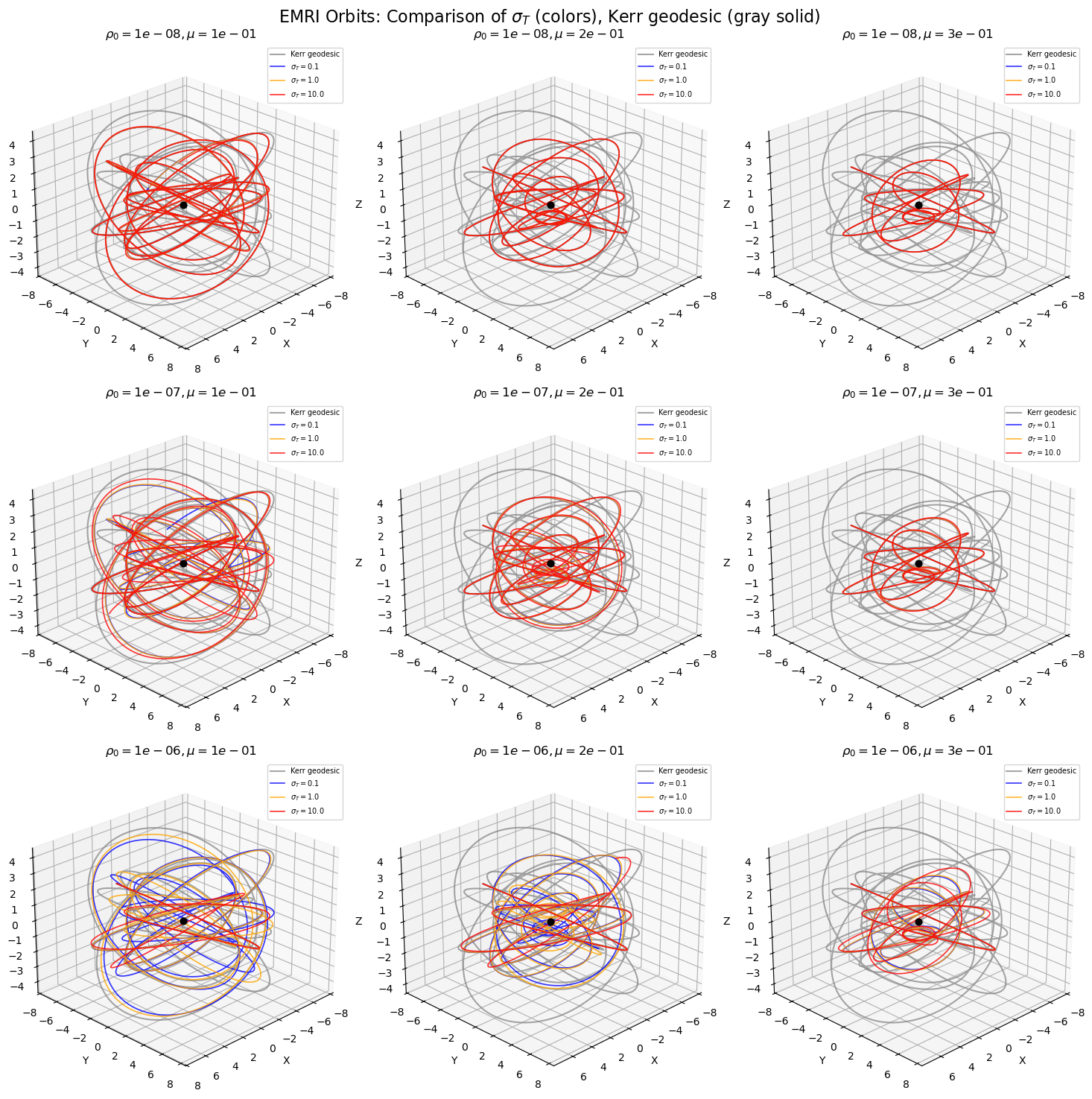}
\caption{
Illustrative three-dimensional orbital trajectories obtained from a
dimensionless parameter scan in geometrized units $G=c=M_{\rm BH}=1$.
The gray curves represent the reference conservative Kerr geodesic, whereas
the colored curves include the dark matter environment, gravitational
radiation reaction, and SIDM-modified dynamical friction. Different rows and
columns illustrate the dependence on the density parameter $\rho_0$ and the
effective dissipative parameter $\mu$, while the colored curves correspond
to different values of the dimensionless self-interaction parameter
$\sigma_T$. The figure is intended to illustrate the qualitative dynamical
trends of the model rather than to represent a specific astrophysical
realization.
}
\label{fig:orbit_scan}
\end{figure}

Figure~\ref{fig:orbit_scan} shows the resulting three-dimensional orbital
trajectories. The gray curves represent the reference conservative Kerr
geodesic, while the colored trajectories include the combined effects of the
dark matter environment, gravitational radiation reaction, and
SIDM-modified dynamical friction. The comparison with the gray curve
therefore illustrates the overall departure from conservative Kerr motion,
rather than isolating a single dissipative contribution.

At fixed $\rho_0$ and $\mu$, the relative differences among the colored
curves provide a direct illustration of the dependence on the effective
self-interaction parameter $\sigma_T$. As $\sigma_T$ is increased, the
additional drag becomes stronger and the orbital evolution departs more
rapidly from the reference trajectory. The effect is also enhanced for
larger values of $\rho_0$, since a denser dark matter environment produces
a stronger dynamical-friction contribution.

The parameter $\mu$ controls the strength of the dissipative evolution in
the present dimensionless calculation. Increasing $\mu$ enhances the
radiation-reaction contribution and leads to a more rapid secular evolution
of the orbit. Consequently, the combined effects of the dark matter
environment and dissipation become increasingly visible across the parameter
scan.

\begin{figure}[htbp]
\centering
\includegraphics[width=0.85\linewidth]{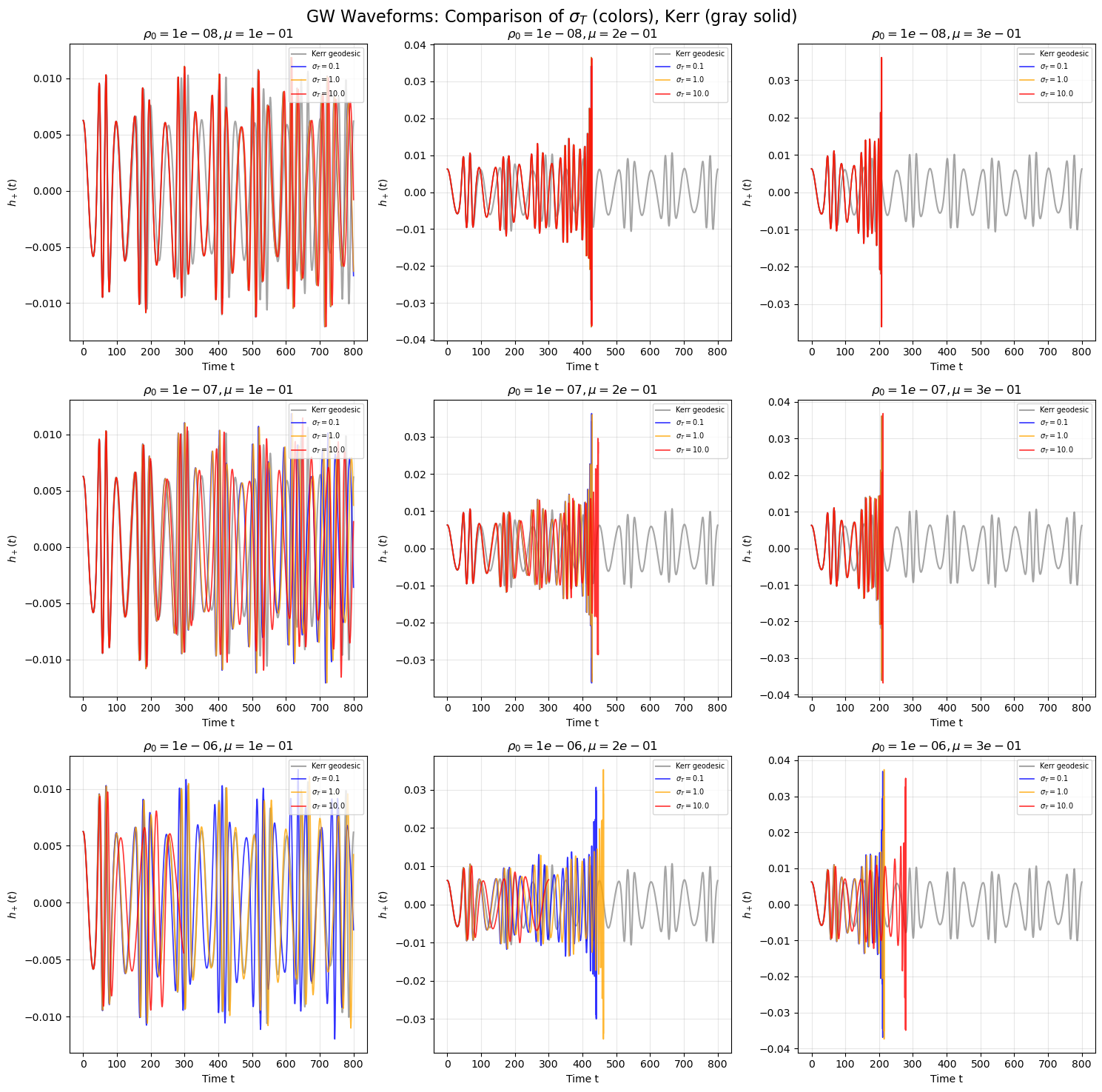}
\caption{
Illustrative gravitational-wave signals corresponding to the dimensionless
orbital evolutions shown in Fig.~\ref{fig:orbit_scan}. The gray curves are
generated from the reference Kerr geodesic, while the colored curves include
the combined effects of the dark matter environment, gravitational radiation
reaction, and SIDM-modified dynamical friction. Increasing the effective
self-interaction parameter leads to progressively larger waveform
differences. The parameter values in this figure are given in dimensionless
code units and are used to demonstrate the qualitative behavior of the
model.
}
\label{fig:waveform_scan}
\end{figure}

The corresponding gravitational-wave signals are shown in
Fig.~\ref{fig:waveform_scan}. The waveform is constructed from the numerical
orbital evolution using a leading-order quadrupole prescription. Relative to
the waveform associated with the reference Kerr trajectory, the
environmentally modified signals gradually develop differences in their
oscillation pattern as the orbital evolution proceeds.

At fixed $\rho_0$ and $\mu$, increasing the effective self-interaction
parameter produces progressively larger waveform modifications. This behavior
is a direct consequence of the enhanced dissipative evolution: changes in
the orbital radius and azimuthal phase modify both the instantaneous
frequency and phase of the associated gravitational-wave signal.

These results provide a qualitative visualization of how environmental
dissipation and dark matter self-interactions can modify EMRI dynamics.
Because several physical effects are included simultaneously in the colored
trajectories, the present figures should not be interpreted as a quantitative
measurement of the isolated SIDM contribution. Instead, they motivate the
frequency-domain analysis in the following section, where the
self-interaction contribution is separated from the collisionless dark matter
effect and the accumulated gravitational-wave phase deviation is evaluated
using physically specified SIDM parameters.

\section{Gravitational-Wave Phase Evolution Induced by SIDM Dynamical Friction}
\label{sec:phase_evolution}

The dynamical friction induced by self-interacting dark matter (SIDM)
provides an additional dissipative channel for extreme-mass-ratio inspirals
(EMRIs). In addition to the orbital energy carried away by gravitational
radiation, the inspiraling compact object transfers energy to the surrounding
dark matter medium through the gravitational wake. The orbital energy balance
can therefore be written as
\begin{equation}
\frac{dE}{dt}
=
\left(\frac{dE}{dt}\right)_{\rm GW}
+
\left(\frac{dE}{dt}\right)_{\rm df},
\end{equation}
where both terms on the right-hand side are negative for an inspiraling orbit.

Using the SIDM-modified dynamical-friction force introduced in the previous
section,
\begin{equation}
\mathbf{F}_{\rm df}^{\rm SIDM}
=
-4\pi G^2 M^2 \rho_\chi(r)
\frac{\ln\Lambda}{v^2}
\left[
1+
\eta
\frac{\rho_\chi(r)}{m_\chi}
\sigma_T
\frac{G M_{\rm BH}}{v^2}
\right]
\hat{\mathbf v},
\end{equation}
the corresponding orbital-energy loss rate is
\begin{equation}
\left(\frac{dE}{dt}\right)_{\rm df}
=
\mathbf{F}_{\rm df}^{\rm SIDM}\cdot\mathbf v,
\end{equation}
which gives
\begin{equation}
\left(\frac{dE}{dt}\right)_{\rm df}
=
-4\pi G^2 M^2\rho_\chi(r)
\frac{\ln\Lambda}{v}
\left[
1+
\eta
\frac{\rho_\chi(r)}{m_\chi}
\sigma_T
\frac{G M_{\rm BH}}{v^2}
\right].
\end{equation}

The first contribution inside the square brackets corresponds to the
conventional collisionless dynamical friction, whereas the second term
represents the leading-order SIDM correction considered in this work.
Within the enhanced-drag branch, $\eta>0$, increasing the
momentum-transfer cross section $\sigma_T$ increases the magnitude of the
additional dissipative contribution.

For a quasi-circular orbit, the orbital frequency $\Omega$ and radius are
related through
\begin{equation}
\Omega^2
=
\frac{G M_{\rm BH}}{r^3}.
\end{equation}
For the dominant quadrupolar gravitational-wave harmonic,
\begin{equation}
f
=
\frac{\Omega}{\pi},
\end{equation}
so that
\begin{equation}
r(f)
=
\left(
\frac{G M_{\rm BH}}{\pi^2 f^2}
\right)^{1/3},
\end{equation}
and
\begin{equation}
v(f)
=
\left(
\pi G M_{\rm BH} f
\right)^{1/3}.
\end{equation}

At leading Newtonian order, the orbital binding energy is
\begin{equation}
E(f)
=
-\frac{1}{2}
M
\left(
G M_{\rm BH}
\right)^{2/3}
\left(
\pi f
\right)^{2/3},
\end{equation}
where the extreme-mass-ratio limit $M\ll M_{\rm BH}$ has been used.
Its frequency derivative is therefore
\begin{equation}
\frac{dE}{df}
=
-\frac{1}{3}
M
\left(
G M_{\rm BH}
\right)^{2/3}
\pi^{2/3}
f^{-1/3}.
\end{equation}

The frequency evolution associated with dynamical friction follows from
\begin{equation}
\left(\frac{df}{dt}\right)_{\rm df}
=
\frac{
(dE/dt)_{\rm df}
}{
dE/df
}.
\end{equation}
Substituting the relations above gives
\begin{equation}
\left(\frac{df}{dt}\right)_{\rm df}
=
\frac{
12 G M \rho_\chi[r(f)]\ln\Lambda
}{
M_{\rm BH}
}
\left[
1+
\eta
\frac{\rho_\chi[r(f)]}{m_\chi}
\sigma_T
r(f)
\right].
\end{equation}

This expression makes the dependence on the SIDM parameters particularly
transparent. The collisionless contribution scales linearly with the local
dark matter density, whereas the leading self-interaction correction contains
the additional factor
\begin{equation}
\frac{\rho_\chi}{m_\chi}\sigma_T r
=
n_\chi\sigma_T r
=
\frac{r}{\lambda_{\rm mfp}}.
\end{equation}

The leading-order vacuum gravitational-wave-driven frequency evolution is
\begin{equation}
\left(\frac{df}{dt}\right)_{\rm GW}
=
\frac{96}{5}
\pi^{8/3}
\left(
\frac{G\mathcal{M}_c}{c^3}
\right)^{5/3}
f^{11/3},
\end{equation}
where $\mathcal{M}_c$ is the chirp mass. The total frequency evolution in
the presence of the dark matter environment is then
\begin{equation}
\dot f_{\rm SIDM}
=
\dot f_{\rm GW}
+
\dot f_{\rm df}^{\rm SIDM}.
\end{equation}

For comparison, the corresponding collisionless-dark-matter evolution is
obtained by setting $\sigma_T=0$,
\begin{equation}
\dot f_{\rm CDM}
=
\dot f_{\rm GW}
+
\dot f_{\rm df}^{(0)},
\end{equation}
whereas the vacuum evolution is simply
\begin{equation}
\dot f_{\rm vac}
=
\dot f_{\rm GW}.
\end{equation}

\subsection{Frequency dependence for a power-law dark matter profile}

To make the frequency dependence of the environmental dissipation explicit,
we consider a power-law dark matter distribution,
\begin{equation}
\rho_\chi(r)
=
\rho_{\rm sp}
\left(
\frac{r_{\rm sp}}{r}
\right)^\alpha,
\end{equation}
where $\rho_{\rm sp}$ and $r_{\rm sp}$ characterize the density normalization
and radial scale, respectively, while $\alpha$ denotes the power-law index.

Using the frequency--radius relation, the density can be written directly as
\begin{equation}
\rho_\chi(f)
=
\rho_{\rm sp}r_{\rm sp}^{\alpha}
\left(
\frac{\pi^2 f^2}{G M_{\rm BH}}
\right)^{\alpha/3}.
\end{equation}

The collisionless dynamical-friction contribution therefore becomes
\begin{equation}
\left(\frac{df}{dt}\right)_{\rm df}^{(0)}
=
A_{\rm df}
f^{2\alpha/3},
\end{equation}
where
\begin{equation}
A_{\rm df}
=
\frac{
12GM\rho_{\rm sp}r_{\rm sp}^{\alpha}\ln\Lambda
}{
M_{\rm BH}
}
\left(
\frac{\pi^2}{GM_{\rm BH}}
\right)^{\alpha/3}.
\end{equation}

The dimensionless SIDM correction can likewise be expressed as
\begin{equation}
n_\chi\sigma_T r
=
\frac{\rho_\chi(r)}{m_\chi}\sigma_T r
=
B_{\rm SI}
f^{-2(1-\alpha)/3},
\end{equation}
where
\begin{equation}
B_{\rm SI}
=
\frac{
\rho_{\rm sp}r_{\rm sp}^{\alpha}\sigma_T
}{
m_\chi
}
\left(
\frac{GM_{\rm BH}}{\pi^2}
\right)^{(1-\alpha)/3}.
\end{equation}

The full SIDM-modified dynamical-friction contribution can thus be written as
\begin{equation}
\dot f_{\rm df}^{\rm SIDM}
=
A_{\rm df}
f^{2\alpha/3}
\left[
1+
\eta B_{\rm SI}
f^{-2(1-\alpha)/3}
\right].
\end{equation}

Equivalently,
\begin{equation}
\dot f_{\rm df}^{\rm SIDM}
=
A_{\rm df}f^{2\alpha/3}
+
\eta A_{\rm df}B_{\rm SI}
f^{(4\alpha-2)/3}.
\end{equation}

The three dissipative mechanisms therefore exhibit distinct frequency
scalings,
\begin{equation}
\dot f_{\rm GW}
\propto
f^{11/3},
\end{equation}
\begin{equation}
\dot f_{\rm df}^{(0)}
\propto
f^{2\alpha/3},
\end{equation}
and
\begin{equation}
\dot f_{\rm SI}
\propto
f^{(4\alpha-2)/3}.
\end{equation}

Consequently, the total SIDM-modified frequency evolution can be written as
\begin{equation}
\dot f_{\rm SIDM}
=
\dot f_{\rm GW}
+
A_{\rm df}
f^{2\alpha/3}
\left[
1+
\eta B_{\rm SI}
f^{-2(1-\alpha)/3}
\right].
\end{equation}

The different power-law dependences imply that the relative importance of
gravitational radiation, collisionless dynamical friction, and SIDM-induced
dynamical friction changes throughout the inspiral. For dark matter profiles
of astrophysical interest, the gravitational-wave radiation term generally
grows more rapidly toward high frequencies, whereas environmental effects are
relatively more important at lower frequencies and larger orbital radii.

The accumulated gravitational-wave phase between a frequency $f$ and a
reference upper frequency $f_{\rm max}$ is
\begin{equation}
\Phi(f)
=
2\pi
\int_f^{f_{\rm max}}
\frac{f'}{\dot f(f')}
\,df'.
\end{equation}

To distinguish the total environmental effect from the additional effect
caused specifically by dark matter self-interactions, we introduce two phase
differences. The total dark-matter-induced dephasing relative to vacuum is
defined as
\begin{equation}
\Delta\Phi_{\rm DM-vac}(f)
=
2\pi
\int_f^{f_{\rm max}}
f'
\left[
\frac{1}{\dot f_{\rm SIDM}(f')}
-
\frac{1}{\dot f_{\rm vac}(f')}
\right]
df'.
\end{equation}

More importantly for the present work, the phase correction generated
specifically by SIDM self-interactions is defined relative to the
collisionless dark matter case,
\begin{equation}
\Delta\Phi_{\rm SIDM-CDM}(f)
=
2\pi
\int_f^{f_{\rm max}}
f'
\left[
\frac{1}{\dot f_{\rm SIDM}(f')}
-
\frac{1}{\dot f_{\rm CDM}(f')}
\right]
df'.
\end{equation}

For the power-law profile introduced above, the latter becomes
\begin{align}
\Delta\Phi_{\rm SIDM-CDM}(f)
=
2\pi
\int_f^{f_{\rm max}}
f'
\Bigg\{
&
\frac{1}{
\dot f_{\rm GW}(f')
+
A_{\rm df}
f'^{\,2\alpha/3}
\left[
1+
\eta B_{\rm SI}
f'^{-2(1-\alpha)/3}
\right]
}
\nonumber\\
&
-
\frac{1}{
\dot f_{\rm GW}(f')
+
A_{\rm df}
f'^{\,2\alpha/3}
}
\Bigg\}
\,df'.
\end{align}

This definition has the useful consistency property
\begin{equation}
\sigma_T\rightarrow0
\quad\Longrightarrow\quad
B_{\rm SI}\rightarrow0
\quad\Longrightarrow\quad
\Delta\Phi_{\rm SIDM-CDM}\rightarrow0.
\end{equation}
In contrast, $\Delta\Phi_{\rm DM-vac}$ remains finite in this limit because
the conventional collisionless dynamical-friction contribution is still
present.

Within the enhanced-drag regime considered here, increasing $\sigma_T$
increases $\dot f_{\rm SIDM}$ and therefore shortens the inspiral timescale.
For the phase convention adopted above, the accumulated SIDM phase is
correspondingly reduced relative to the collisionless-dark-matter case, so
that $\Delta\Phi_{\rm SIDM-CDM}$ is generally negative. Its absolute value,
$|\Delta\Phi_{\rm SIDM-CDM}|$, increases with the self-interaction strength
and provides a convenient measure of the cumulative waveform modification.

Since EMRIs remain in the observational band for a large number of orbital
cycles, even relatively small instantaneous corrections to the frequency
evolution can accumulate into substantial phase differences. The resulting
dephasing therefore provides a potentially sensitive probe of SIDM effects
in the dark matter environment surrounding massive black holes.

\section{Numerical Results and Discussion}
\label{sec:results}

We numerically evaluate the gravitational-wave phase evolution for an
extreme-mass-ratio inspiral (EMRI) consisting of a $5M_\odot$ compact object
orbiting a $10^5M_\odot$ massive black hole. The dark matter density
normalization is fixed to
\begin{equation}
\rho_{\rm max}
=
5\times10^{-7}\,{\rm kg\,m^{-3}},
\end{equation}
and we adopt $\ln\Lambda=3$. The dark matter particle mass and the effective
response coefficient are fixed to
\begin{equation}
m_\chi=10^3\,{\rm GeV}/c^2,
\qquad
\eta=0.5.
\end{equation}

To investigate the dependence on the microscopic self-interaction strength,
we consider four representative values of the momentum-transfer cross section,
\begin{equation}
\sigma_T
=
0,\quad
10^{-28}\,{\rm m^2},\quad
10^{-27}\,{\rm m^2},\quad
10^{-26}\,{\rm m^2}.
\end{equation}
The case $\sigma_T=0$ corresponds to the collisionless dark matter limit,
for which the additional SIDM correction vanishes while the conventional
Chandrasekhar dynamical-friction contribution remains present.

\begin{figure}[t]
\centering
\includegraphics[width=0.55\textwidth]{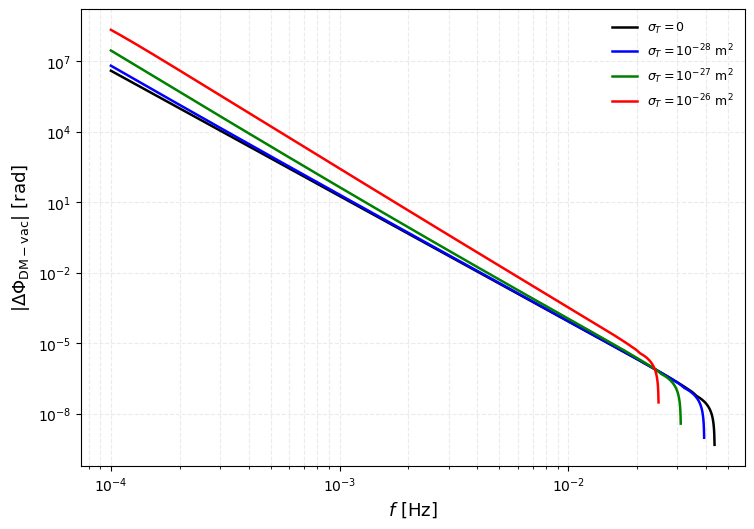}
\caption{
Absolute gravitational-wave phase difference
$|\Delta\Phi_{\rm DM-vac}|$ relative to the vacuum inspiral for different
momentum-transfer cross sections. The $\sigma_T=0$ curve corresponds to
collisionless dark matter, while the nonzero-$\sigma_T$ curves additionally
include the SIDM-induced correction to dynamical friction. The total
environmental phase deviation increases with the self-interaction strength.
}
\label{fig:sidm_phase_dm_vac}
\end{figure}

We first consider the total environmental phase modification relative to the
vacuum inspiral,
\begin{equation}
\Delta\Phi_{\rm DM-vac}
=
\Phi_{\rm SIDM}
-
\Phi_{\rm vac}.
\end{equation}
The corresponding absolute phase difference,
$|\Delta\Phi_{\rm DM-vac}|$, is shown in
Fig.~\ref{fig:sidm_phase_dm_vac}. The collisionless case,
$\sigma_T=0$, already produces a nonzero phase deviation because the compact
object experiences conventional dynamical friction from the surrounding dark
matter distribution. Increasing $\sigma_T$ introduces the additional SIDM
correction and progressively increases the total environmental dephasing.

Although the total phase difference is useful for quantifying the overall
environmental effect, it contains contributions from both the collisionless
dark matter halo and the additional self-interaction correction. To isolate
the latter, we define
\begin{equation}
\Delta\Phi_{\rm SIDM-CDM}
=
\Phi_{\rm SIDM}
-
\Phi_{\rm CDM},
\end{equation}
where $\Phi_{\rm CDM}$ denotes the phase obtained in the same dark matter
environment after setting $\sigma_T=0$. Since the conventional
Chandrasekhar contribution is present in both phases, this subtraction
directly isolates the additional modification generated by SIDM
self-scattering.

\begin{figure}[t]
\centering
\includegraphics[width=0.55\textwidth]{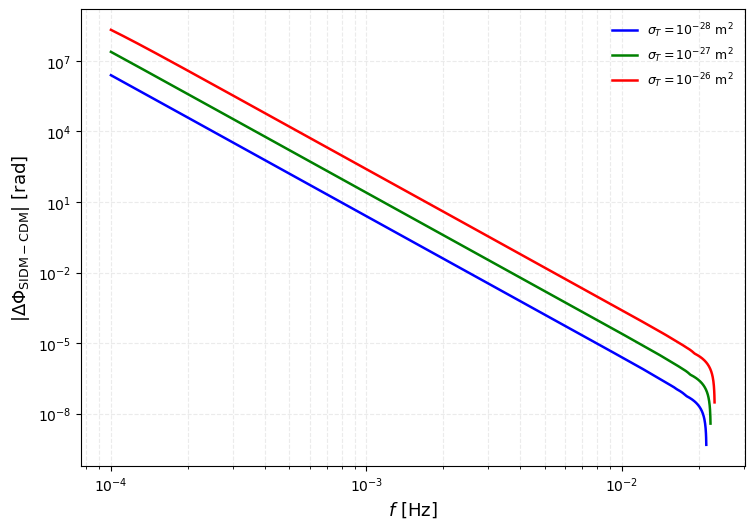}
\caption{
Absolute SIDM-induced gravitational-wave phase difference
$|\Delta\Phi_{\rm SIDM-CDM}|$ relative to the collisionless dark matter
case. The collisionless contribution common to both models has been
subtracted, so that the curves directly represent the additional phase
modification associated with dark matter self-interactions. Increasing
$\sigma_T$ produces a progressively larger accumulated dephasing.
}
\label{fig:sidm_phase_sidm_cdm}
\end{figure}

The resulting quantity $|\Delta\Phi_{\rm SIDM-CDM}|$ is shown in
Fig.~\ref{fig:sidm_phase_sidm_cdm}. By construction,
\begin{equation}
\sigma_T\rightarrow0
\qquad\Longrightarrow\qquad
\Delta\Phi_{\rm SIDM-CDM}\rightarrow0.
\end{equation}
The SIDM-induced dephasing increases systematically with the momentum-transfer
cross section. This monotonic behavior reflects the leading-order dependence
of the effective dynamical-friction correction on $\sigma_T$.

Representative numerical values are listed in
Table~\ref{tab:sidm_phase}. At $f\simeq10^{-3}\,{\rm Hz}$, the additional
SIDM-induced phase difference is approximately $2.5$ rad for
$\sigma_T=10^{-28}\,{\rm m^2}$, while it increases to about $25$ rad and
$251$ rad for $\sigma_T=10^{-27}\,{\rm m^2}$ and
$10^{-26}\,{\rm m^2}$, respectively. The approximately linear dependence on
$\sigma_T$ in this frequency range is consistent with the leading-order
effective correction adopted in this work.

\begin{table}[t]
\centering
\caption{
Representative absolute gravitational-wave phase differences for different
momentum-transfer cross sections. The quantity
$|\Delta\Phi_{\rm DM-vac}|$ measures the total dark matter environmental
effect relative to vacuum, whereas $|\Delta\Phi_{\rm SIDM-CDM}|$ isolates
the additional contribution associated specifically with dark matter
self-interactions.
}
\label{tab:sidm_phase}
\begin{tabular}{c|cc|cc|cc}
\hline
$\sigma_T$ &
\multicolumn{2}{c|}{$f=10^{-4}\,{\rm Hz}$} &
\multicolumn{2}{c|}{$f=10^{-3}\,{\rm Hz}$} &
\multicolumn{2}{c}{$f=10^{-2}\,{\rm Hz}$} \\
$(\mathrm{m^2})$ &
$|\Delta\Phi_{\rm DM-vac}|$ &
$|\Delta\Phi_{\rm SIDM-CDM}|$ &
$|\Delta\Phi_{\rm DM-vac}|$ &
$|\Delta\Phi_{\rm SIDM-CDM}|$ &
$|\Delta\Phi_{\rm DM-vac}|$ &
$|\Delta\Phi_{\rm SIDM-CDM}|$ \\
\hline
$0$
& $3.90\times10^{6}$ & $0$
& $1.82\times10^{1}$ & $0$
& $8.42\times10^{-5}$ & $0$ \\

$10^{-28}$
& $6.38\times10^{6}$ & $2.48\times10^{6}$
& $2.07\times10^{1}$ & $2.51$
& $8.67\times10^{-5}$ & $2.48\times10^{-6}$ \\

$10^{-27}$
& $2.83\times10^{7}$ & $2.44\times10^{7}$
& $4.33\times10^{1}$ & $2.51\times10^{1}$
& $1.09\times10^{-4}$ & $2.48\times10^{-5}$ \\

$10^{-26}$
& $2.15\times10^{8}$ & $2.11\times10^{8}$
& $2.69\times10^{2}$ & $2.51\times10^{2}$
& $3.33\times10^{-4}$ & $2.49\times10^{-4}$ \\
\hline
\end{tabular}
\end{table}

The strong frequency dependence of the phase shift is a consequence of the
competition between environmental dissipation and gravitational-wave
radiation reaction. At low frequencies, the compact object remains in the
inspiral for a comparatively long time and accumulates a large number of
orbital cycles. In addition, the characteristic collisionality parameter,
\begin{equation}
N_{\rm scat}
=
n_\chi\sigma_T r,
\end{equation}
increases toward larger orbital radii. These effects combine to produce a
large accumulated dephasing at the lowest frequencies considered.

At higher frequencies, gravitational-wave radiation reaction becomes
increasingly dominant. Consequently, both the collisionless dark matter
effect and the additional SIDM contribution decrease rapidly as the compact
object approaches the late inspiral. This behavior is clearly visible in
Figs.~\ref{fig:sidm_phase_dm_vac} and
\ref{fig:sidm_phase_sidm_cdm}.

The numerical collisionality diagnostics also provide an estimate of the
domain of validity of the leading-order effective description. Over the full
frequency range considered here, the maximum values of $N_{\rm scat}$ are
approximately
\begin{equation}
N_{\rm scat}^{\rm max}
\simeq
1.44,\quad
14.4,\quad
144
\end{equation}
for
\begin{equation}
\sigma_T
=
10^{-28},\quad
10^{-27},\quad
10^{-26}\,{\rm m^2},
\end{equation}
respectively. The larger-cross-section cases therefore extend beyond the
strict weak-collision condition $N_{\rm scat}\ll1$ at the lowest frequencies
and should be interpreted as illustrative extrapolations of the leading-order
model in that region.

A more conservative regime occurs around
$f\sim10^{-3}\,{\rm Hz}$ for the smaller self-interaction cross sections.
Importantly, the additional SIDM contribution can already reach the
order-radian level in this region. In particular,
\begin{equation}
|\Delta\Phi_{\rm SIDM-CDM}|
\simeq
2.5\ {\rm rad}
\end{equation}
for $\sigma_T=10^{-28}\,{\rm m^2}$ at
$f\simeq10^{-3}\,{\rm Hz}$.

Previous studies of EMRI signals have commonly used a gravitational-wave
dephasing of order $0.1$ rad as a rough benchmark for phase resolvability by
LISA for signal-to-noise ratios of order $20$--$30$
\cite{Gupta2021,Maselli2022}. The SIDM-induced phase difference obtained here
is therefore well above this commonly adopted benchmark, suggesting that the
effect may be potentially resolvable in favorable observational scenarios.
A definitive assessment of detectability, however, requires a full
waveform-overlap or parameter-estimation analysis.

This result illustrates an important property of EMRIs: even a modest
modification of the instantaneous orbital decay rate can accumulate over many
cycles into a substantial gravitational-wave phase difference.

The numerical behavior can be understood directly from the effective
SIDM-modified dynamical-friction force,
\begin{equation}
F_{\rm df}^{\rm SIDM}
=
4\pi G^2 M^2\rho_\chi(r)
\frac{\ln\Lambda}{v^2}
\left[
1+
\eta
\frac{\rho_\chi(r)}{m_\chi}
\sigma_T
\frac{G M_{\rm BH}}{v^2}
\right],
\end{equation}
where the expression denotes the magnitude of the drag force. Within the
enhanced-drag branch considered here, the additional term grows linearly
with $\sigma_T$ at leading order. Therefore,
\begin{equation}
\sigma_T
\uparrow
\quad\Longrightarrow\quad
|F_{\rm df}^{\rm SIDM}|
\uparrow
\quad\Longrightarrow\quad
\dot f_{\rm SIDM}
\uparrow
\quad\Longrightarrow\quad
t_{\rm insp}
\downarrow,
\end{equation}
which ultimately leads to an increasing accumulated phase deviation.

It is also important to distinguish the roles of the macroscopic dark matter
distribution and the microscopic self-interaction strength. The density
$\rho_\chi$ determines the amount of dark matter participating in the
gravitational wake and controls the conventional Chandrasekhar contribution.
In contrast, $\sigma_T$ characterizes momentum transfer among dark matter
particles and enters the additional SIDM correction. The two quantities
therefore influence the inspiral through physically distinct but coupled
mechanisms.

The large number of orbital cycles accumulated during an EMRI substantially
enhances the sensitivity of the waveform to weak environmental effects. The
quantity $\Delta\Phi_{\rm SIDM-CDM}$ is particularly useful because it removes
the ordinary collisionless dark matter contribution and directly isolates
the additional phase modification generated by self-interactions.

The present analysis adopts a leading-order effective description motivated
by the collisional SIDM framework and focuses on the enhanced-drag branch.
A more complete treatment would require a self-consistent calculation of the
collisional gravitational wake, including the velocity dependence of
$\sigma_T$, possible modifications of the dark matter phase-space
distribution, and relativistic corrections to the orbital dynamics.
A quantitative assessment of observational detectability would additionally
require dedicated parameter-estimation studies with realistic detector noise
models.

\section{Conclusions}
\label{sec:conclusions}

In this work, we investigated the influence of self-interacting dark matter
(SIDM) on the orbital dynamics and gravitational-wave signatures of
extreme-mass-ratio inspirals (EMRIs) around massive black holes. Starting
from the collisional Boltzmann description of SIDM, we identified the
dimensionless collisionality parameter
\begin{equation}
n_\chi \sigma_T r
=
\frac{r}{\lambda_{\rm mfp}},
\end{equation}
which measures the importance of dark matter self-scattering on the
characteristic scale of the gravitational wake. Motivated by this kinetic
description, we constructed a leading-order effective correction to the
standard Chandrasekhar dynamical-friction force.

The resulting dynamical-friction model directly connects the microscopic
momentum-transfer cross section $\sigma_T$ with the macroscopic dissipative
force acting on the inspiraling compact object. In the enhanced-drag branch
considered in this work, the effective force takes the form
\begin{equation}
\mathbf F_{\rm df}^{\rm SIDM}
=
\mathbf F_{\rm df}^{(0)}
\left[
1+
\eta n_\chi\sigma_T r
\right],
\end{equation}
where $\eta$ characterizes the effective response of the gravitational wake
to dark matter self-scattering. Although the coefficient $\eta$ is not
derived from a complete solution of the full collisional Boltzmann equation,
the dependence on $n_\chi\sigma_T r$ follows naturally from the scattering
rate and mean-free-path scales of the SIDM medium.

Using this effective dynamical-friction prescription, we numerically studied
the EMRI orbital evolution and the associated gravitational-wave signals. Our
results show that, within the enhanced-drag regime, increasing the SIDM
self-interaction strength increases the magnitude of the dynamical friction
and accelerates the inspiral relative to the collisionless dark matter case.
The effect becomes progressively stronger as the momentum-transfer cross
section increases, reflecting the enhanced environmental dissipation induced
by dark matter self-scattering.

We further calculated the corresponding gravitational-wave frequency and
phase evolution. For a power-law dark matter distribution, the collisionless
and SIDM-induced contributions to the frequency evolution exhibit distinct
power-law dependences on the gravitational-wave frequency. This difference
allows the environmental contribution to accumulate over the long-duration
EMRI inspiral and produce a substantial phase deviation.

To separate the overall dark matter effect from the additional contribution
generated specifically by self-interactions, we introduced two phase
differences: the total dark-matter-induced dephasing relative to vacuum,
$\Delta\Phi_{\rm DM-vac}$, and the additional SIDM-induced dephasing relative
to the collisionless dark matter case, $\Delta\Phi_{\rm SIDM-CDM}$. By
construction,
\begin{equation}
\sigma_T\rightarrow0
\qquad\Longrightarrow\qquad
\Delta\Phi_{\rm SIDM-CDM}\rightarrow0,
\end{equation}
so that $\Delta\Phi_{\rm SIDM-CDM}$ directly isolates the phase modification
associated with dark matter self-interactions.

The numerical results demonstrate that the magnitude of the accumulated phase
difference increases with the momentum-transfer cross section in the
enhanced-drag regime. Since EMRIs spend a large number of orbital cycles in
the observational band, even relatively small instantaneous modifications of
the orbital decay rate can accumulate into significant waveform dephasings.
This makes the gravitational-wave phase evolution particularly sensitive to
environmental dissipative effects.

Our results therefore demonstrate the potential of EMRI gravitational waves
to probe dark matter self-interactions in the vicinity of massive black
holes. The framework developed here provides a direct connection between
microscopic SIDM scattering properties, the dynamical friction experienced by
the inspiraling compact object, and observable modifications of the
gravitational-wave phase evolution. Future space-based detectors such as LISA
may consequently provide complementary information on SIDM models beyond
that obtained from conventional astrophysical systems.

The present analysis adopts a leading-order effective treatment of the SIDM
correction and focuses on the enhanced-drag branch. A more complete
description would require a self-consistent calculation of the collisional
gravitational wake from the full kinetic equation, including the detailed
velocity dependence of the self-interaction cross section, possible
modifications of the dark matter phase-space distribution, and relativistic
corrections to the orbital dynamics. Combining such an improved dynamical
model with dedicated parameter-estimation analyses will be an important
direction for future work.

\section*{acknowledgements}
The authors gratefully acknowledge support from the project ``Integrated Electronics Technology for Inertial Sensors'' (2024YFC2207003) led by Meilin Liu. 

This work is also supported by the China--Brazil Belt and Road Joint Laboratory on Radio Astronomy Technology, and the National Key Research and Development Program of China under the project ``Strategic Science and Technology Innovation Cooperation'' (No. 2025YFE0212600), which is also led by Meilin Liu.

The authors also thank all colleagues who provided helpful discussions and assistance during this work, and especially Professor Haiguang Xu for his guidance and support.

\section*{Funding}
This work is supported by the National Key Research and Development Program of China (Nos. 2024YFC2207003, 2025YFE0212600) and related programs under the China--Brazil Belt and Road Joint Laboratory on Radio Astronomy Technology.

\section*{Data Availability Statement}
This manuscript has no associated data. 
[Author’s comment: Data sharing is not applicable to this article as no datasets were generated or analyzed during the current study.]

\section*{Code Availability Statement}

Code/software will be made available on reasonable request.

[Author’s comment: The code/software generated during and/or analyzed during the current study is available from the corresponding author on reasonable request.]


\begin{thebibliography}{99}

\bibitem{AmaroSeoane2017}
P.~Amaro-Seoane et al.,
``Laser Interferometer Space Antenna,''
arXiv:1702.00786.

\bibitem{BarackCutler2004}
L.~Barack and C.~Cutler,
``LISA capture sources: Approximate waveforms, signal-to-noise ratios,
and parameter estimation accuracy,''
Phys. Rev. D \textbf{69}, 082005 (2004),
doi:10.1103/PhysRevD.69.082005.

\bibitem{AmaroSeoane2007}
P.~Amaro-Seoane, J.~R.~Gair, M.~Freitag, M.~C.~Miller,
I.~Mandel, C.~J.~Cutler and S.~Babak,
``Intermediate and extreme mass-ratio inspirals:
astrophysics, science applications and detection using LISA,''
Class. Quantum Grav. \textbf{24}, R113--R169 (2007),
doi:10.1088/0264-9381/24/17/R01.

\bibitem{Gair2013}
J.~R.~Gair et al.,
``Testing general relativity with low-frequency, space-based
gravitational-wave detectors,''
Living Rev. Relativ. \textbf{16}, 7 (2013).

\bibitem{Barausse2014}
E.~Barausse, V.~Cardoso and P.~Pani,
``Can environmental effects spoil precision gravitational-wave astrophysics?,''
Phys. Rev. D \textbf{89}, 104059 (2014),
doi:10.1103/PhysRevD.89.104059.

\bibitem{Planck2018}
N.~Aghanim et al. (Planck Collaboration),
``Planck 2018 results. VI. Cosmological parameters,''
Astron. Astrophys. \textbf{641}, A6 (2020).

\bibitem{Merritt2010}
D.~Merritt,
``Dynamics and Evolution of Galactic Nuclei,''
Princeton University Press (2010).

\bibitem{GondoloSilk1999}
P.~Gondolo and J.~Silk,
``Dark matter annihilation at the galactic center,''
Phys. Rev. Lett. \textbf{83}, 1719 (1999).

\bibitem{Chandrasekhar1943}
S.~Chandrasekhar,
``Dynamical friction. I. General considerations,''
Astrophys. J. \textbf{97}, 255 (1943).

\bibitem{Eda2013}
K.~Eda et al.,
``New probe of dark matter distribution in the Galactic center using
gravitational wave signals,''
Phys. Rev. D \textbf{88}, 023006 (2013).

\bibitem{Macedo2013}
C.~F.~B.~Macedo, P.~Pani, V.~Cardoso and L.~C.~B.~Crispino,
``Into the lair: gravitational-wave signatures of dark matter,''
Astrophys. J. \textbf{774}, 48 (2013),
doi:10.1088/0004-637X/774/1/48.

\bibitem{Cardoso2022}
V.~Cardoso, K.~Destounis, F.~Duque, R.~Panosso Macedo and A.~Maselli,
``Gravitational waves from extreme-mass-ratio systems in astrophysical
environments,''
Phys. Rev. Lett. \textbf{129}, 241103 (2022),
doi:10.1103/PhysRevLett.129.241103.

\bibitem{Nichols2023}
D.~A.~Nichols, B.~A.~Wade and A.~M.~Grant,
``Secondary accretion of dark matter in intermediate mass-ratio inspirals:
Dark-matter dynamics and gravitational-wave phase,''
Phys. Rev. D \textbf{108}, 124062 (2023),
doi:10.1103/PhysRevD.108.124062.

\bibitem{BinneyTremaine2008}
J.~Binney and S.~Tremaine,
``Galactic Dynamics,''
2nd ed., Princeton University Press (2008).

\bibitem{Navarro1997}
J.~F.~Navarro et al.,
``A universal density profile from hierarchical clustering,''
Astrophys. J. \textbf{490}, 493 (1997).

\bibitem{Einasto1965}
J.~Einasto,
``On the construction of a composite model for the Galaxy,''
Trudy Inst. Astrofiz. Alma-Ata \textbf{5}, 87 (1965).

\bibitem{Wang2025CQG}
Y.~Wang, W.-B.~Han, X.~Wu and E.~Liang,
``Gravitational wave imprints of dark matter annihilation:
probing halo structures through EMRI signals,''
Class. Quantum Grav. \textbf{42}, 175007 (2025),
doi:10.1088/1361-6382/adf7fb.

\bibitem{Wang2025SIDM}
Y.~Wang, R.~Tang, W.-B.~Han and E.~Liang,
``Distinguishing Strongly Interacting Dark Matter Spikes via EMRI
Gravitational Waves,''
Symmetry \textbf{17}, 1878 (2025),
doi:10.3390/sym17111878.

\bibitem{Kavanagh2017}
B.~J.~Kavanagh et al.,
``Probing dark matter with gravitational waves,''
Phys. Rev. D \textbf{96}, 103013 (2017).

\bibitem{Spergel2000}
D.~N.~Spergel and P.~J.~Steinhardt,
``Observational evidence for self-interacting cold dark matter,''
Phys. Rev. Lett. \textbf{84}, 3760 (2000).

\bibitem{Feng2010}
J.~L.~Feng, M.~Kaplinghat and H.-B.~Yu,
``Sommerfeld enhancements for thermal relic dark matter,''
Phys. Rev. D \textbf{82}, 083525 (2010),
doi:10.1103/PhysRevD.82.083525.

\bibitem{BuckleyFox2010}
M.~R.~Buckley and P.~J.~Fox,
``Dark matter self-interactions and light force carriers,''
Phys. Rev. D \textbf{81}, 083522 (2010),
doi:10.1103/PhysRevD.81.083522.


\bibitem{Petraki2014}
K.~Petraki, L.~Pearce and A.~Kusenko,
``Self-interacting asymmetric dark matter coupled to a light massive
dark photon,''
JCAP \textbf{07}, 039 (2014),
doi:10.1088/1475-7516/2014/07/039.

\bibitem{DelNobile2015}
E.~Del Nobile, M.~Kaplinghat and H.-B.~Yu,
``Direct detection signatures of self-interacting dark matter with a
light mediator,''
JCAP \textbf{10}, 055 (2015),
doi:10.1088/1475-7516/2015/10/055.

\bibitem{Chu2020}
X.~Chu, C.~Garcia-Cely and H.~Murayama,
``A practical and consistent parametrization of dark matter
self-interactions,''
JCAP \textbf{06}, 043 (2020),
doi:10.1088/1475-7516/2020/06/043.

\bibitem{TulinYu2018}
S.~Tulin and H.-B.~Yu,
``Dark matter self-interactions and small scale structure,''
Phys. Rep. \textbf{730}, 1 (2018).

\bibitem{Kaplinghat2016}
M.~Kaplinghat et al.,
``Dark matter cores and cusps in galaxies,''
Phys. Rev. Lett. \textbf{116}, 041302 (2016).

\bibitem{Vogelsberger2012}
M.~Vogelsberger et al.,
``Subhalos in self-interacting dark matter cosmology,''
Mon. Not. Roy. Astron. Soc. \textbf{423}, 3740 (2012).

\bibitem{Rocha2013}
M.~Rocha et al.,
``Cosmological simulations with self-interacting dark matter I,''
Mon. Not. Roy. Astron. Soc. \textbf{430}, 81 (2013).

\bibitem{FischerSagunski2024}
M.~S.~Fischer and L.~Sagunski,
``Dynamical friction from self-interacting dark matter,''
Astron. Astrophys. \textbf{690}, A299 (2024),
doi:10.1051/0004-6361/202451304.

\bibitem{Boudon2024}
A.~Boudon, P.~Brax, P.~Valageas and L.~K.~Wong,
``Gravitational waves from binary black holes in a self-interacting
scalar dark matter cloud,''
Phys. Rev. D \textbf{109}, 043504 (2024),
doi:10.1103/PhysRevD.109.043504.

\bibitem{AlonsoAlvarez2024}
G.~Alonso-\'Alvarez, J.~M.~Cline and C.~Dewar,
``Self-Interacting Dark Matter Solves the Final Parsec Problem of
Supermassive Black Hole Mergers,''
Phys. Rev. Lett. \textbf{133}, 021401 (2024),
doi:10.1103/PhysRevLett.133.021401.

\bibitem{Berti2015}
E.~Berti et al.,
``Testing general relativity with present and future astrophysical
observations,''
Class. Quantum Grav. \textbf{32}, 243001 (2015),
doi:10.1088/0264-9381/32/24/243001.

\bibitem{Gupta2021}
P.~Gupta, B.~Bonga, A.~J.~K.~Chua and T.~Tanaka,
``Importance of tidal resonances in extreme-mass-ratio inspirals,''
Phys. Rev. D \textbf{104}, 044056 (2021),
doi:10.1103/PhysRevD.104.044056.

\bibitem{Maselli2022}
A.~Maselli, N.~Franchini, L.~Gualtieri, T.~P.~Sotiriou,
S.~Barsanti and P.~Pani,
``Detecting fundamental fields with LISA observations of gravitational
waves from extreme mass-ratio inspirals,''
Nature Astron. \textbf{6}, 464--470 (2022),
doi:10.1038/s41550-021-01589-5.

\end{thebibliography}
\end{document}